# Online News Media Website Ranking Using User Generated Content

Samaneh Karimi[1,2], Azadeh Shakery[1,3] and Rakesh Verma[2]

[1]School of Electrical and Computer Engineering, College of Engineering, University of Tehran, Tehran, Iran
[2]Computer Science Department, University of Houston, Houston, USA
[3]Institute for Research in Fundamental Sciences (IPM), Tehran, Iran,
samanekarimi@ut.ac.ir, shakery@ut.ac.ir, rverma@uh.edu

## Abstract

News media websites are important online resources that have drawn great attention of text mining researchers. The main aim of this study is to propose a framework for ranking online news websites from different viewpoints. The ranking of news websites is useful information, which can benefit many news-related tasks such as news retrieval and news recommendation. In the proposed framework, the ranking of news websites is obtained by calculating three measures introduced in the paper and based on user-generated content. Each proposed measure is concerned with the performance of news websites from a particular viewpoint including the completeness of news reports, the diversity of events being covered by the website and its speed. The use of user-generated content in this framework, as a partly-unbiased, real-time and low cost content on the web distinguishes the proposed news website ranking framework from the literature. The results obtained for three prominent news websites, BBC, CNN, NYTimes, show that BBC has the best performance in terms of news completeness and speed, and NYTimes has the best diversity in comparison with the other two websites.

## 1. Introduction

In recent years, the wide growth of news content on the web, and the considerable portion (11%) of news-related queries entered by users into search engines [1], has attracted the information science research community's attention to the online news area and its challenges. One of the main online news providers is news media websites, or briefly news websites, which constantly compete with each other in reporting events and breaking news. Therefore, ranking them based on aspects that help users choose between different news sources is becoming more important.

After a significant event, in addition to news reports published by news websites, a surge of event-related content is generated by ordinary users through personal blogs, wikis and social networks. Twitter is one of the well-known social networks widely used by people as a source of news on events. According to a study conducted by Pew research center [2], in 2013, 52% of American Twitter users used it as a source for news and this share increased to 63% in 2015 [3]. Due to the short length of tweets, most events are reflected promptly in Twitter and disseminated quickly. So, Twitter content can act as an up-to-date online content. Moreover, tweets are generated by a huge number of ordinary users (not only professional content producers), so Twitter content may be less prone to be affected by a particular party, or group of biased news content producers, especially in reporting events. Thus, the provision of information about events, which is freely available, makes Twitter a low-cost source of online information about the news. In the literature, this kind of content is described as user-generated content (UGC), participatory journalism, and citizen journalism [4].

In this paper, a framework for ranking news websites based on user-generated content is proposed. In addition to the mentioned benefits of UGC as news-related online content such as being up-to-date, low-cost and partly-unbiased, it also provides useful information for analyzing the performance of news websites. For example, the ability of a news website to cover most (or all) of the noteworthy events in each time period without missing the important ones is intuitively considered as a positive characteristic. This characteristic can be considered as a metric for ranking news websites. To this aim, UGC can be used as a rich source of information related to noteworthy events in the intended time period including the number of events, the estimated time of occurrence of each event and the description of each event.

In this framework, some performance measures, which are specifically defined to analyze news websites' performance and rank them, are introduced. News diversity is a measure proposed to quantify the performance of news website in covering as many events as possible in each time period. A higher news diversity value indicates that the corresponding news website performs better in satisfying the subset of readers who seek for diverse news by reporting more events. The next proposed measure is news completeness which shows how detailed and complete the news articles are in reporting the events. The third proposed measure is speed which indicates how fast the news websites are in reporting the events since their time of occurrence. The importance of speed for news websites in the world of intense competition for reporting the news first is obvious. Knowing the values of these measure helps readers in choosing their favorite news website according to their needs. In addition to the proposed measures, a search engine-based website ranking method is proposed to obtain another ranking of the websites regarding the same set of events and to analyze the ranking of the news websites using two different methods.

The website evaluation problem has been addressed by researchers during the last few years. Many methods have been proposed for evaluating different types of websites from different aspects such as design, content, accessibility, credibility and usability of the website [5]. One of the differences between this paper and some of the previous works is the focus of this paper on a specific type of website, i.e. news websites, and utilizing this specificity in the proposed approach. Another difference of this paper and related papers in website evaluation is that the main goal in this paper is to facilitate comparing news websites from different viewpoints by defining new measures. Different viewpoints of comparing news websites include different quality measures defined to evaluate the performance of the news website. It is worth mentioning that these measures are not supposed to show the absolute superiority of one website to another as most papers in website evaluation area aim to do. The proposed framework tries to provide the basis for analyzing news websites from different points of view and benefit other news-related tasks with its output. The framework's output can be interpreted and employed differently in accordance with the application that it is used for.

In many previous works, websites are evaluated through a survey [5-13] in which participants respond to questions about different aspects of website performance and the final evaluation is based on their responses. This approach is called empirical approach in the literature. Using human participants in the evaluation process has some limitations including:
- The result is vulnerable to be affected by subjective information received from the limited number of survey participants and there is a chance of obtaining biased results.
- The employment of respondents could be costly.
- The limited number of survey participants in empirical methods [14] narrows the utility of the results.

In this paper, a different approach for news website ranking is chosen to eliminate the mentioned problems. Since the focus of this paper is on news websites, the basis of the framework is the set of events detected from user generated content and information that can be extracted from this kind of content. As events are detected from a large number of tweets generated by users, the "small number of participants" problem would be alleviated. In the proposed approach, the similarity of news website's articles to the detected events based on different aspects, such as content, diversity of reported events or their publish times, leads to higher scores for the news website in the corresponding aspects.

Another challenge of questionnaire-based methods is providing appropriate questions so that the required information can be elicited from the participants/experts, while we only need to provide a collection of tweets. Then, the proposed framework automatically detects the events from tweets and calculates the measures based on extracted events. More detailed comparison between this work and previous works is given in the related work section. Overall, the following research questions are studied in this paper:
- What are the important performance aspects for ranking news websites (RQ1)?
- How news websites' performance aspects can be quantified through defining and automatically computing some metrics (RQ2)?
- Can user generated content such as tweets be used for ranking news websites based on news websites' content quality (RQ3)?

Therefore, the main contributions of the paper are briefly summarized as follows:
- Proposing a framework for ranking news websites, which is based on events detected from tweets as user generated content.
- Proposing news-specific ranking measures using the language modeling approach.
- Proposing a search engine-based website ranking method for a known set of events.

The rest of this paper is organized as follows. Section 2 reviews some related works. Section 3 describes the proposed framework. Evaluation of the proposed framework and experimental results are presented in Section 4. Conclusions are given in Section 5.

## 2. Related Work

There are several previous methods for analyzing and evaluating the performance of websites from different domains [5,7-13,15-19]. These methods mostly evaluate the website as an interface and study different aspects such as website's design, content, accessibility, and usability [5] and have been used for various domains such as e-commerce websites [13], e-government websites [82], agritourism websites [21], tourism management [22], library websites [23] and news websites [5,24-26]. It is important to clarify the position of this work in the literature, express its similarities with previous works and point out its relative advantages.

### 2.1. Position of this study in the literature

In the literature, the notion of website evaluation is used for different tasks with different goals. In [19], two criteria for describing websites are defined: the content of the website which refers to "what is presented?" and its design which is related to the question "how is it presented?" Our proposed method concerns with the quality of content published by online news media. In other words, all of the measures introduced in this paper investigate what is presented in online news websites. So, this work can be categorized into the first group of studies that investigate "what is presented?".

The authors in [27] classify usability evaluation methods into two types: empirical methods, which basically analyze usage data from real end-users, and inspection methods, which are based on reviewing the usability aspects by expert evaluators. In another paper, [17], website usability evaluation methods are categorized into four groups: automatic (software evaluation), empirical (user testing), formal (evaluation models), and informal (expert evaluation). Based on this categorization, this paper is most similar to the automatic methods category. The main similarity between previous automatic methods and our method is that no respondents, either normal users or experts, are employed in both methods so the evaluation is not based on the respondents' responses to a questionnaire or their feedback while using the website. Therefore, in the next section, previous works on website evaluation that are closely related to our research are reviewed in two groups, based on the employment of human responses in the evaluation: automatic methods [14, 16, 26, 28-30] and empirical methods [5, 20, 24, 25, 83].

## 2.2. Website Evaluation

### 2.2.1. Automatic Methods

According to [31], an automatic website evaluation method is a "software that automates the collection of interface usage data and identifies potential Web problems." In [30], the authors have proposed a method that automatically determines the comprehensibility level of a website using a linear regression model and six categories of quantitatively computed features: Information Value, Information Credibility, the Media Instructional Value, Affective Attention rating, Organization Structure and Usability of the website.

Zhitomirsky-Geffet et al. [26] proposed a method for evaluating the reliability and bias of the written media on the web. Their method is based on comparing the textual content of news websites' articles to political texts with known tendency. To this end, a classifier is learnt using a labeled dataset of political texts and then used to tag news websites' articles. It is also shown that readers' perception of bias is highly correlated with the classifier's results.

In [16], the author listed some web-testing tools and discussed how automatic website testing tools can affect the quality assurance processes to improve production and maintenance of websites. Rodríguez et al. [29] proposed an automatic method for evaluating county e-Governance maturity level based on analysis of municipal websites. This method employs automatic tools such as W3C validator, Xeno s/w and source code analyzer. A comprehensive study on usability evaluation is published by Ivory and Hearst [14], where various approaches for automating usability testing methods are discussed. These approaches include a capture technique to log user activity automatically and an analysis of log files made by Web servers to record client requests.

Most previous works in automatic website evaluation try to automate evaluating the user interface, whereas this paper focuses on automatic analysis of the content published by news websites. Moreover, these tools detect a number of issues that mostly have to be later inspected by humans [16], while the proposed method does not require direct judgment from human respondents. It ranks news websites by extracting the required event related information from tweets published at the same time.

### 2.2.2. Empirical Methods

The empirical category is the most often used category among the four groups of website evaluation methods [32, 33]. In the study published by Al-Radaideh et al. [5], the authors empirically evaluated the Jordanian online newspapers with respect to the website usability and content factors based on users' point of view. Usability of Malaysian online news websites is evaluated in [25] with a questionnaire-based approach. The study published by Li [24] employed an empirical approach to analyze three U.S. online newspapers' contents comparing the use of textual and graphical elements.

Besides website evaluation, other topics have been also studied using empirical methods. As an example, the combined impacts of on-air (TV) and online network (Web) news, as two kinds of media, on student and adult perceptions of news credibility is investigated in [6]. In a study published by Cassidy [20], journalists' perceptions of online news credibility are assessed through a survey of 655 US online and print journalists. The results of their survey show that "online news is perceived as moderately credible overall." Since the present work is focused on online news websites, the next section is devoted to a review of the studies on this topic.

## 2.3. News Websites Evaluation

Chung, Nam and Stefanone [7] focused on the credibility factors of three different categories of online news sources: mainstream (e.g., usatoday.com), independent (e.g., thedrudgereport.com) and index-type websites (e.g., news.google.com). The authors investigate the impact of technological characteristics of online news including interactivity, multimediality, and hypertextuality on their credibility assessments using a questionnaire-based approach. Hypertextuality refers to "the ability to connect within-sites and otherwise disparate webpages through clicking on a word, phrase, or graphic image," [7].

The use of different news sources on news consumers' credibility perceptions is also studied using an empirical approach. Kruikemeier and Lecheler [11], consider three main categories of news sources in their study: social media, online media, and traditional sourcing techniques such as calling a source or going to a press conference. The authors design a vignette study to test the impact of visible verification of a journalist source on the credibility of news sources and comparison between traditional and new online sources. In a vignette study, participants are asked to provide their evaluation of different systems [11]. The authors show that social media, online media and traditional sourcing techniques are perceived as the least credible, moderately credible, and credible respectively.

One of the most similar papers in the literature to the present work, in terms of using comments as a kind of user generated content for assessment of news articles credibility, is proposed by Pjesivac, Geidner and Cameron [18] who try to test the impact of source expertise and opinion valence[i] in reader's comments embedded in a news story about genetically modified organisms (GMOs) on the credibility of the news story. The results of their questionnaire-based study show that the expertise level of the comment writer has an important impact on the perceived credibility of the news article. This result supports the previous works' finding, which indicated that "Internet users mainly use the peripheral or heuristic route of information processing to evaluate online news credibility." Besides, they show that the opinion valence expressed in the comments does not have any significant impact on the credibility assessment.

In [4], the relationship between consumption, creation and perception of news is examined. In this study, the results of reliability tests from a survey of US adults show that the role of news consumption is more significant than news creation in predicting attitudes toward citizen journalism and the professional tenets of good journalism. They also compare the general news consumers with citizen journalism consumers and show that they have different views of the professional tenets of good journalism.

Reviewing the studies in website evaluation domain, we see that evaluating the performance of websites from the content quality point of view and independent of a group of human participators' information has not widely been investigated. Besides, most of the studies that address the problem of news websites evaluation define some indicators or employ some criteria developed by other papers for evaluation and analyze the websites' usefulness from users' perspective. The main weakness of most existing methods that use evaluation criteria is that the criteria are generally based on individual author's opinions and preferences [34]. In the present work, a novel approach for analyzing the news websites' performance is proposed that is not based on intuitively-defined criteria but is based on events automatically detected from user generated content and how news articles published by the news websites report the events. In other words, in the proposed method, the content generated by the crowd indicate what should be reported, how should it be reported, and when should it be reported. To achieve this aim, two components including an event detection component and a news website ranking component are introduced in the proposed framework. In the website ranking component, some metrics are proposed that facilitate comparing news websites from different viewpoints. The proposed metrics are not necessarily aimed to value the websites and label them as good or bad, they are defined in order to extract the information about how news websites perform with regard to the events found in Twitter. This information can be beneficial in many news-related tasks such as news retrieval and news recommendation. As an example, the news websites rankings can be used to rerank their published documents retrieved in response to a query, thereby incorporating the news ranking measures in news retrieval score computations. News website rankings can also be utilized in news recommendation for customizing the article suggestions based on its users' preferences and their favorable news content characteristics. So, in the rest of this section, the papers that study the impact of tweets on news reporting and journalism along with the methods that employ tweets for news-related tasks including event detection and news recommendation are reviewed.

## 2.4. Tweets in News Area

With the advent and rapid growth of Twitter, a number of studies have focused on investigating the impact of user generated content such as tweets on news-related topics, e.g., news resource selection and journalistic routines [35-40,81]. In [37], different studies that investigate the use of Twitter in gathering, reporting and disseminating the news are reviewed. Lecheler and Kruikemeier [39] study the impact of online sources on selection of news resources by journalists and also the strategies employed to verify a news resource. Twitter is mentioned as an important resource for journalism researchers in [39], but mainly for soft news and when no other suitable information resource is available. The use of tweets by journalists is also studied by Broersma and Graham [36], where the authors investigated how journalists use tweets as quotes in news articles, the sections of the article in which tweets are included and what kinds of sources are quoted, through an analysis on eight British and Dutch national newspapers over a five-year period.

Using the Twitter stream for event detection has recently become a popular area of research. Event detection has been studied in the long-running Topic Detection and Tracking (TDT) [41] research program. Some of the event detection papers propose methods for finding events using news articles [42-50], others focus on event detection directly from tweets [51-58, 88, 89] and a third group uses both news articles and tweets [59-63]. Event detection methods can be categorized into two approaches: document-pivot approach [64-66, 49] and feature-pivot approach [67-69]. In document-pivot approaches, documents are clustered, then event-related features are extracted from each cluster. In feature-pivot approach, features that are useful for event detection are identified and extracted from the stream, then events are detected by clustering these features [56].

Event detection methods and their common traits and differences have been studied in several papers [70, 84-86]. Garg and Kumar [70] survey a wide range of event detection techniques from clustering and categorization to topic modeling approaches. Furthermore, multivariate research on event detection, such as employing weather data, temporal data and geo-location data, is discussed. Another survey on event detection methods applied to streaming Twitter data is conducted by Hasan et al. [84].

In addition to event detection, tweets are used for other news related tasks such as news recommendation or news summarization. In [59], the authors proposed a user modeling framework that constructs three different kinds of user profiles for news recommendation. The user profiles are hashtag-based, entity-based and topic-based that are constructed using hashtags, entities and topics mentioned in tweets and their related news articles. In their paper, tweets are employed to construct user profiles for news recommendation, while we use tweets for event detection and analyzing news websites in terms of their event reporting performance. In [90], an abstractive summarization method is proposed by employing interaction mechanisms to provide hierarchically ordered summaries of news tweets on a certain event along a time line. The interaction mechanisms used in [90] include presenting information in a bullet-style summary in addition to conception expansion that allows the user to view complementary information about each concept.

Recommending tweets for news articles is another related task. In [71], the authors used 16 Twitter features such as publication time, length and follower count plus the similarity scores of tweets and news articles to find the most relevant tweets to each news article. Connecting news articles to Twitter conversations has also been studied in previous works [72, 87, 93]. For instance, in the framework proposed by Shi, Ifrim and Hurley [72] the tweets are separated per article based on keyword similarity. Then, four features: local cosine similarity, global cosine similarity, local frequency of the hashtag and global frequency of the hashtag, are extracted for each article-hashtag pair and used for classification.

To summarize, tweets have been found useful in many news related tasks. On the other hand, most existing news websites evaluation approaches employ intuitively defined criteria that lack a theoretical background. Therefore, employing tweets instead, which is generated by a wide range of users, for analyzing the performance of news websites and ranking them can be beneficial.

## 3. Methodology

The task of website evaluation can be viewed from different perspectives that range from analyzing the quality of content published by the website to interface usability assessment approaches. In this paper, a framework for analyzing and ranking news websites is proposed, which leverages the resources of the news area to define appropriate measures for ranking (RQ1). For this purpose, Twitter, as an online resource, which is highly pertinent to news area, is employed to study the third research question (RQ3). Figure 1 shows the overall schema of the proposed news website ranking framework.

The idea of the framework is to detect events using user generated content such as tweets and then compute the ranking measures for news websites based on the detected events and the news articles published by them. In other words, the news websites' performance is analyzed by ranking them according to their scores on the proposed measures. The measures consider three aspects of a news website performance: diversity, reporting different kinds of events, speed in publishing news articles that report an event, and providing in-depth and comprehensive articles that cover important details of the corresponding event.

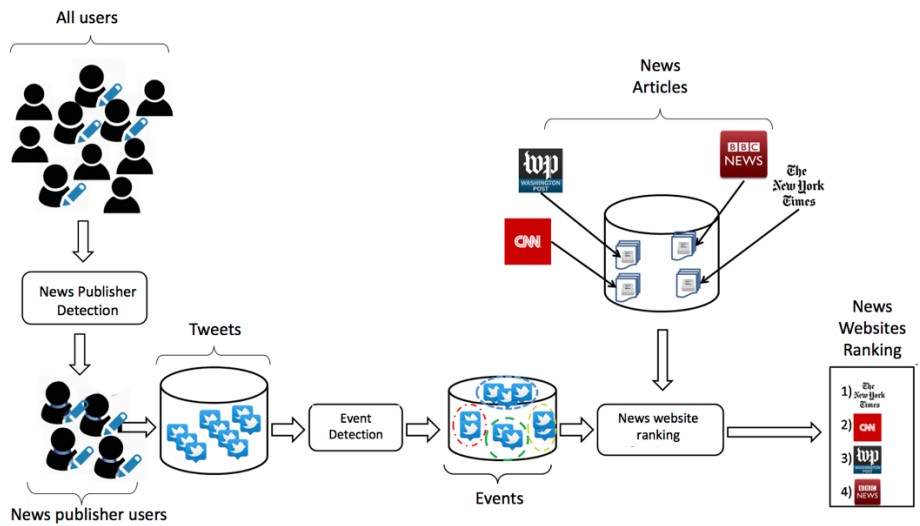

Figure 1. The overall schema of the proposed news website ranking framework.

As shown in Figure 1, the framework consists of two main components: event detection and news website ranking. The inputs of the framework are tweets written by users and the concomitant news articles published by news websites. The output of the framework is a ranking of the news websites, which represents their performance on that time period according to the proposed measures. In the rest of this section, each component is explained in detail.

## 3.1. Event Detection Component

The aim of this component is to find events during a specific time period based on tweets during this period. The reason for employing tweets to detect events are the special characteristics of Twitter. One Twitter characteristic is fast event dissemination, events are reported by its users immediately after (sometimes even during) the occurrence, as the message length limitation in Twitter means that tweets are short and can be written quickly. Meanwhile, other longer messages or documents like news articles of the news websites need more time to be written, edited and posted. This characteristic makes Twitter messages an up-to-date source of events, and the earliest publish time of messages on an event is usually a good estimate of the event's occurrence time. Another important characteristic of tweets is that they are written by a large population of ordinary people, so the messages are less likely to be biased by specific policies, considerations or rules, which can be the case with some news websites and their news articles. Although, note that the Twitter content can also be affected by propagandists or government agencies that try to publish biased data using their Twitter accounts. The detection of this type of biased content is beyond the scope of this paper.

In general, the topics of tweets are very diverse and can be relevant or irrelevant to the news and events so excluding the news-irrelevant tweets can improve the performance of the event detection method and consequently the proposed news websites ranking method. We use a news publisher detection method introduced by [53] in the framework to classify Twitter users into 'news publisher' and 'others.'

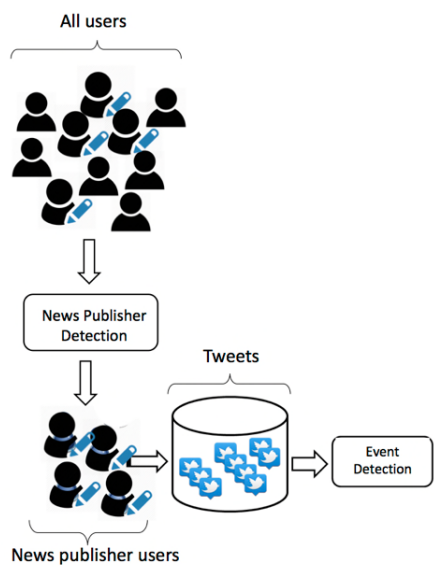

Figure 2. Event detection component.

In the news publisher detection step, two sets of features are used according to [53]. Briefly, the first set of features represents metadata information about twitter users: number of followers, number of following, ratio of number of followers to number of following, number of tweets that the user posted, number of tweets the user favorited, number of lists the user follows and two Boolean features indicating that the user account is verified or not and the user profile contains a URL or not. The second set of features represents the terms in the username and description of the user. In this paper, the second set of features are represented as unigram presence features. Then, a Bayesian network classifier [73] implemented in WEKA [74] is employed to classify the twitter users to 'news publisher' or 'others.'

After the news publisher detection step, tweets written by users who are classified as 'news publisher' are used as inputs of the event detection method. We use a short text clustering method called GSDMM, a collapsed Gibbs Sampling algorithm for the Dirichlet Multinomial Mixture model for short text clustering [75], for event detection. The main idea of this method is to repeatedly compute the probability that document d is generated by cluster z based on the words of the document and the language model of the clusters until the probabilities converge.

The output of the event detection component provides useful information for the framework to analyze the performance of the news websites such as the set of events, a basic representative for each event and an approximation of each event's occurrence time. The set of events detected by GSDMM are used as inputs of the news website ranking component. In the rest of this section, this component is explained in more detail.

## 3.2. News Website Ranking Component

As our aim is to analyze the performance of news websites, any news related quality can be employed to define efficient and informative measures to analyze the performance. To achieve this aim, three performance measures are defined based on the events detected by the first component and the news articles published by the websites. In the rest of this section, performance measures defined to analyze the websites' performance are introduced (RQ2).

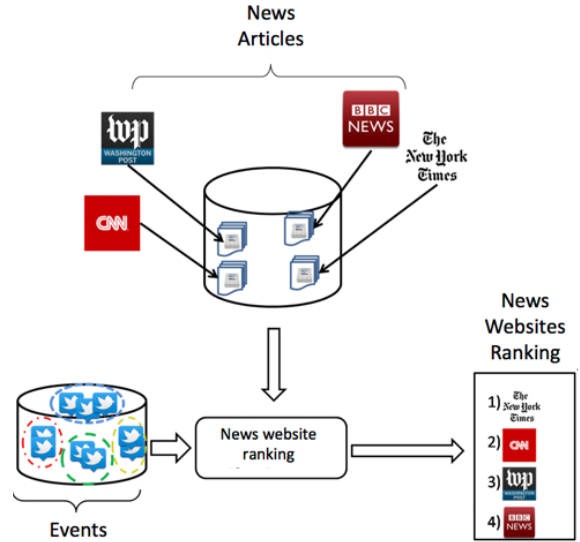

Figure 3. Website ranking component.

### 3.2.1. News Diversity

At each time period, a set of events attracts people's attention and makes them look for news resources to know more. One of the measures that reflects an important aspect of a news website's performance is the number of events covered by the website in each time period. In order to calculate this measure, we use tweets generated by users at a time period to represent each event occurred at that time period. To analyze the performance of the website $S_i$ in covering different events, the News_Diversity measure is introduced as follows:

$$E: \{e_1, e_2, \ldots, e_k\}$$

$$A_i: \{a_1, a_2, \ldots, a_p\}$$

$$\text{News\_Diversity}(S_i, T) = \frac{|\{e_j \mid \exists j \in (1,k), \exists l \in (1,p) ; \text{Rel}(a_l, e_j) = 1\}|}{k}$$

$$\text{Rel}(a_l, e_j) = \begin{cases} 1 & \text{if news article } a_l \text{ is reporting the event } e_j \\ 0 & \text{if news article } a_l \text{ is not reporting the event } e_j \end{cases}$$

where E shows the set of events detected by the first component and occurred at time period T, $A_i$ shows the set of news articles published by website $S_i$ at time period T, k is the total number of events occurred at time period T and p is the number of news articles published by website $S_i$. The value of function Rel(.) which shows whether a news article is reporting an event or not is based on the Jensen-Shannon divergence [76] similarity value between each news article's language model and each tweet cluster's language model. Each news article is assigned to its most similar tweet cluster, so the value of function Rel($a_l$, $e_j$) would be one if event $e_j$ is the most similar event to news article $a_l$, based on their Jensen-Shannon divergence similarity value.

### 3.2.2. News Completeness

One of the important aspects of a news website's performance is how complete and detailed the events are reported by the news website [5, 20]. Therefore, in the second measure we try to score the completeness of news articles that are reporting an event. In this measure, we employ the output of the first component, which is event detection from tweets, as a representative for the event and compute the similarity between each event and the news articles published by the website to report the event. In other words, we consider the completeness of the event representation extracted from tweets to be 100% and try to estimate the completeness of news articles in reporting the same event. Other event detection methods or sources of information can also be employed for representing the events in this measure, however the event representation used in this paper facilitates benefiting from user generated content, as mentioned in the introduction, and language model representation. The language model "provides a principled way to quantify the uncertainties associated with the use of natural language," [91]. In this measure, the language model of each news article, $\theta_{a_l}$, and the language model of its corresponding event, $\theta_{e_{a_l}}$, are compared using the Jensen-Shannon divergence score. The language model of news article $a_l$, which is $\theta_{a_l}$, using the maximum likelihood estimation is computed as:

$$\theta_{a_l}: \{P_{ML}(w_i|\theta_{a_l}) = \frac{c(w_i, a_l)}{\sum_{w \in V_{a_l}} c(w, a_l)}\}_{i=1}^{|V_{a_l}|}$$

where $V_{a_l}$ is the vocabulary of words in document $a_l$ and $c(w_i, a_l)$ is the frequency of word $w_i$ in document $a_l$. In order to estimate the language model of each event, all of the tweets in the tweets cluster are considered as one document so the language model of event $e_{a_l}$, which is $\theta_{e_{a_l}}$, associated with the news article $a_l$ using the maximum likelihood estimation is computed as:

$$\theta_{e_{a_l}}: \{P_{ML}(w_i|\theta_{e_{a_l}}) = \frac{c(w_i, e_{a_l})}{\sum_{w \in V_{e_{a_l}}} c(w, e_{a_l})}\}_{i=1}^{|V_{e_{a_l}}|}$$

where $e_{a_l}$ represents the event that news article $a_l$ is reporting, $V_{e_{a_l}}$ shows the vocabulary of words in tweet cluster $e_{a_l}$ and $c(w_i, e_{a_l})$ is the frequency of word $w_i$ in tweet cluster $e_{a_l}$.

To compute the News_Completeness measure for website $S_i$, the similarity between each news article $a_l$ that is published by $S_i$ and its corresponding event $e_{a_l}$, is computed using Jensen-Shannon divergence as follows:

$$D(\theta_{a_l} || \theta_{e_{a_l}}) = \sum_{w \in V} P(w|\theta_{a_l}) \log \frac{P(w|\theta_{a_l})}{P(w|\theta_{e_{a_l}})}$$

$$\text{JS\_divergence}(\theta_{a_l}, \theta_{e_{a_l}}) = \frac{D(\theta_{a_l}||\theta_{e_{a_l}}) + D(\theta_{e_{a_l}}||\theta_{a_l})}{2}$$

$$\text{JS\_SimilarityScore}(\theta_{a_l}, \theta_{e_{a_l}}) = - \text{JS\_divergence}(\theta_{a_l}, \theta_{e_{a_l}})$$

Then, News_Completeness measure is computed by finding the average over all similarity values between the pairs of news articles and their corresponding event at time period T:

$$\text{News\_Completeness}(S_i, T) = \frac{\sum_{a_l \in A'_i} \text{JS\_SimilarityScore}(\theta_{a_l}, \theta_{e_{a_l}})}{|A'_i|}$$

where $A'_i$ represents the set of event reporting news articles published by the website $S_i$.

### 3.2.3. Speed

The next measure examines how fast a website acts in reporting events. To compute this measure, we need to have an estimation of the occurrence time of the event. The time of the event can be estimated using the times that tweets representing that event are written:

$$\text{Time}(e) = \min_{t \in \text{Tweets}(e)}(\text{Time}(t))$$

where Tweets(e) is the set of tweets forming the cluster of event e and Time(t) is the publication time of tweet t. Therefore, to compute the speed measure for news website $S_i$ at time period T, we have:

$$\Delta = |\text{EarliestNewsArticlePublishTime} - \text{SecondEarliestNewsArticlePublishTime}|$$

$$\text{Speed}(S_i, T) = \text{average}_{e \in E_{S_i}} \left( \frac{\max_{S_j \in S^e}\left(\text{Time}(a^e_{S_j})\right) - \min_{S_j \in S^e}\left(\text{Time}(a^e_{S_j})\right) + \Delta}{\left(\text{Time}(a^e_{S_i})\right) - \min_{S_j \in S^e}\left(\text{Time}(a^e_{S_j})\right) + \Delta} \right)$$

where $E_{S_i}$ represents the set of events reported by news website $S_i$, $S^e$ represents the set of news websites reporting event e, $a^e_{S_i}$ represents the earliest news article published by $S_i$ reporting event e and $\Delta$ is the interval between the publish time of the earliest and the second earliest news articles reporting event e, among all websites' news articles that are reporting event e. In case only one news article is reporting event e, $\Delta$ is set to a small constant value. This value is used to prevent having a zero-value denominator. The speed measure computes the ratio of the interval between the slowest and the fastest websites' publish time of their news articles reporting event e and the same interval between website $S_i$ and the slowest website for event e. A low value of the denominator implies a fast news website and high speed value.

### 3.3. Comparison with Similar Measures

Table 1 summarizes all of the proposed measures and a sample of measures used by scholars related to evaluation of news websites [5, 7, 20]. As shown in Table 1, some of these measures calculate similar aspects of news websites performance but the measures proposed in our paper to analyze the websites' performance have two particular characteristics: 1) they are specifically defined for news websites' analysis 2) the analysis is fully automatic. For example, the news completeness measure proposed in our paper and Context/Coverage measure used in [5] and comprehensiveness measure used in [20] evaluate similar aspects of news websites, the difference is that the news completeness measure is calculated automatically based on user generated content and the other two measures are computed through a survey.

The timeliness measure [92] is an automatically computed measure that evaluates the same performance aspect as the speed measure proposed in our paper. Both timeliness and speed measures score news websites based on their articles publish times in reporting events, however there are subtle differences in their formulations. To the best of our knowledge, there is no measure introduced by previous works, similar to the news diversity measure introduced in our paper.

Table 1. Website performance measures proposed in this paper and related works.

| Measure | Description |
| --- | --- |
| Multimediality [7] | To what extent text, graphics, and (moving) images with sound are translated and integrated into a common digital form |
| Context/Coverage [5] | The depth or breadth of the information provided on the website |
| Fairness [20] | How fair is the news information from Internet sources |
| Comprehensiveness [20] | How comprehensive is the news information from Internet sources |
| Timeliness [92] | How promptly a news stream publishes |
| News diversity (The present work) | How many events have been covered by the news website |
| News completeness (The present work) | How completely the events are reported in news articles |
| Speed (The present work) | How fast a website acts in reporting events |

# 4. Results and Discussion

## 4.1. Datasets

To evaluate the performance of our proposed methods, we used a dataset of tweets and a dataset of news articles released in [60] that are both generated within the same time period and are available on the supporting website of the corresponding paper [77]. The tweet dataset contains more than 2,000,000 tweets that are crawled from Twitter information streams of 1,425 users over a period of more than two months (starting from the end of October 2010 to the beginning of January 2011). The news article dataset contains 77,544 news articles monitored from news websites including BBC, CNN and New York Times [60].

For the proposed news publisher detection method, a training dataset is needed so we used a dataset of Twitter users' information employed in [53]. This dataset contains metadata and textual features for 10,000 Twitter users. In this dataset, 157 users are labeled as 'news publisher' and the rest of them are labeled as 'others.' We also collected the same set of user information features as our training data for Twitter users of the tweet dataset.

## 4.2. First Component Evaluation

### 4.2.1. News Publisher Detection

As explained in Section 3.1, the first step in the proposed framework is detecting the news publisher users in Twitter. To implement this step, two sets of features are computed for training and testing data according to [53]. The first set includes eight metadata features about the Twitter user and the second set includes textual features, which are the users' usernames and description terms represented as unigram presence features. We used OpenNLP tokenizer (http://opennlp.apache.org/) and Stanford Lemmatizer [78] as our preprocessing tools to tokenize and lemmatize the usernames and description sections. Finally, we achieved 7,024 textual features for train data and employed them along with 8 metadata features to train a Bayes Network classifier [79] using Weka tool [74].

The classifier labeled 27 users out of 1,425 users of the test set as 'news publisher' and the rest of them are labeled as 'others.' Table 2 shows usernames and descriptions of the users labeled as 'news publisher.'

Table 2. Username and description of users labeled as 'news publisher'

| Number | Username | Description |
|---|---|---|
| 1 | KrisKetzKMBC | @KMBC / @KCWE_TV Weeknight Anchor. 32 years at KMBC. (Links & RT's are not endorsements. Opinions are mine.) Husband to the wonderful @RadioDana on |
| | | @KMBZRadio |
| 2 | CBS21NEWS | Your station for news, sports and weather in Central Pennsylvania. |
| 3 | Jeannie_Hartley | Unprecedented change we bring by navigating a course uncharted #Equality will exist when all #HumanRights are honored globally @HumansFoRights focus: #Solutions |
| 4 | CNNTravel | Your insider guide to the latest travel news and the worlds' best travel, entertainment and lifestyle experiences. |
| 5 | alaskaHQ | Social journalist/author/broadcaster, now proudly @ https://t.co/J8oKqOC0WM, First Look Media. Also @attila_the_cat's mom. On Egypt since #Jan25. #FreeAlaa. |
| 6 | AtlantaNewsFeed | Atlanta News on Twitter |
| 7 | jane__bradley | 'Another investigative journalist who thinks they work for MI5.' Investigations Correspondent @BuzzfeedUK. Formerly of BBC/Panorama. PGP: http://t.co/OAYxGrvUO0 |
| 8 | CNNRadio | Listen to CNN on SiriusXM, TuneInRadio and with podcasts in iTunes or your favorite podcast app. |
| 9 | Omid_M | Journalist, #Iran Analyst, Grad of UC Berkeley's #Journalism School, #Politics, #MiddleEast, #Art, & Soccer enthusiast |
| 10 | LisaFranceCNN | Senior producer for CNN Digital's Entertainment section. Passionate about too many things. Only Judy can judge me. RTs are not endorsements |
| 11 | grattongirl | Best-selling author & global social media strategist. In the 'Sunday Times Social List', & in 'Twitter's Top 75 Badass Women' #BA75 (Wife of @grattonboy) |
| 12 | carmensoo | null |
| 13 | FGoria | Managing Editor @CatchyBigData. Editor-in-Chief @EAST_ec. Finance geek @Corriereit. SMM @ScuolaGraffer. Mountaineer. Trail runner. CAI and AAC member. |
| 14 | followfriday200 | I am the http://t.co/vpfrLgsjZt Bot, I will tweet the top ranked tweeps. You don't need to follow me, YOU CAN |
| | | FOLLOW @followfridaycom INSTEAD |
| 15 | SantaTeresaNews | The place for news of the Santa Teresa & Sunland Park, New Mexico area. |
| 16 | ProphetDLYoung | Twitter for Prophet Derrick Young Follow my business @DyEnterprises |
| 17 | grattonboy | Dean (hubby to @grattongirl) is a superb cook, foodie, wine lover, bestselling author and columnist for @skypeinmedia. @ADVAOpticalNews and @telecoms. |
| 18 | showbiztonight | A.J. Hammer hosts @ShowbizTonight Monday - Thursday at 11pm ET/PT on @HLNTV http://t.co/2WhGV58A5U |
| 19 | CSRwire | The latest #news, views and reports in #CSR and #sustainability. Part of the @3BLMedia Group. |
| 20 | JoshEstrin | Pop Culture fascinates me... |
| 21 | HHDukeMehal | The Council on Foreign Relations is dedicated to one world government, and with preserving and defending the U.S. constitution and our free-enterprise system. |
| 22 | harrisj | Innovation Specialist at @18F, proud former troublemaker at @nytimes. Data scold. Personal account: @harrisj_self |
| 23 | BM_AG | Birmingham Museum & Art Gallery (BMAG), a world class museum in the heart of Birmingham city centre. Part of Birmingham Museums Trust. |
| 24 | topresonancers | I am the Resonancers Bot, I will notify you whenever your tweets resonate.   You don't need to follow me, YOU CAN FOLLOW @resonancers INSTEAD. |
| 25 | RobertMackey | Senior Writer at @the_intercept. Previously, @nytimes reporter, @NYTOpenSource columnist and editor of @thelede blog. |
| 26 | WGNNews | Chicago's very own source for news, weather, sports and entertainment. Join us on Facebook http://t.co/SDUdw1qFP4 & Instagram https://t.co/PIWgjjG87Y |
| 27 | MarsGroupKenya | All human beings are born Free and Equal in Dignity and Rights. |

As shown in Table 2, most of the users in the test set that are detected as news publishers by our method are indeed news media Twitter accounts and some of them are Twitter accounts of journalists or users who actively report events or news related contents. By checking the accounts of users in Table 2 and considering their description by an expert, 20 users are recognized to be correctly classified as news publisher so the precision would be:

$$\text{Precision} = \frac{\text{number of users correctly classified as news publisher}}{\text{total number of users classified as new publisher}} = \frac{20}{27} = 0.74$$

The 20 news publishers report news in different domains such as sports, e.g. user #1, travel, e.g. user #4, finance, e.g. user #13 or technology, e.g. user #11 and #17. The numbers of 7 users that are incorrectly classified as news publisher in Table 2 are #12, #14, #16, #20, #21, #23 and #24.
The main aim of this step is purifying the tweet dataset by excluding irrelevant tweets. Also, we believe that any mistakes here would roughly equally handicap/help all news websites.

The next step in the proposed framework is detecting events using tweets written by news publisher users. The news publisher step reduced the tweet dataset size from 2,316,204 tweets to 64,685 tweets which are tweets written by 27 users labeled as news publisher. The obtained set of tweets is then preprocessed by OpenNLP tokenizer and Stanford Lemmatizer and some tokens such as URLs or '@mentions' (terms starting with '@') in addition to stop words are eliminated. Next, the preprocessed set of tweets is used by GSDMM method to find the events. In this paper, the values of GSDMM parameters are $\alpha = 0.1$, $\beta = 0.1$, K = 50 and the number of iterations is 10.

GSDMM method found 50 clusters of tweets with different sizes. Figure 4 shows the word cloud representation for some of the clusters found by GSDMM method (cluster numbers 9, 14, 16, 20 and 24). In this Figure, the size of the words in each cluster is proportional to their weights in the corresponding cluster. The word clouds are generated using an online tool called WordItOut (https://worditout.com).

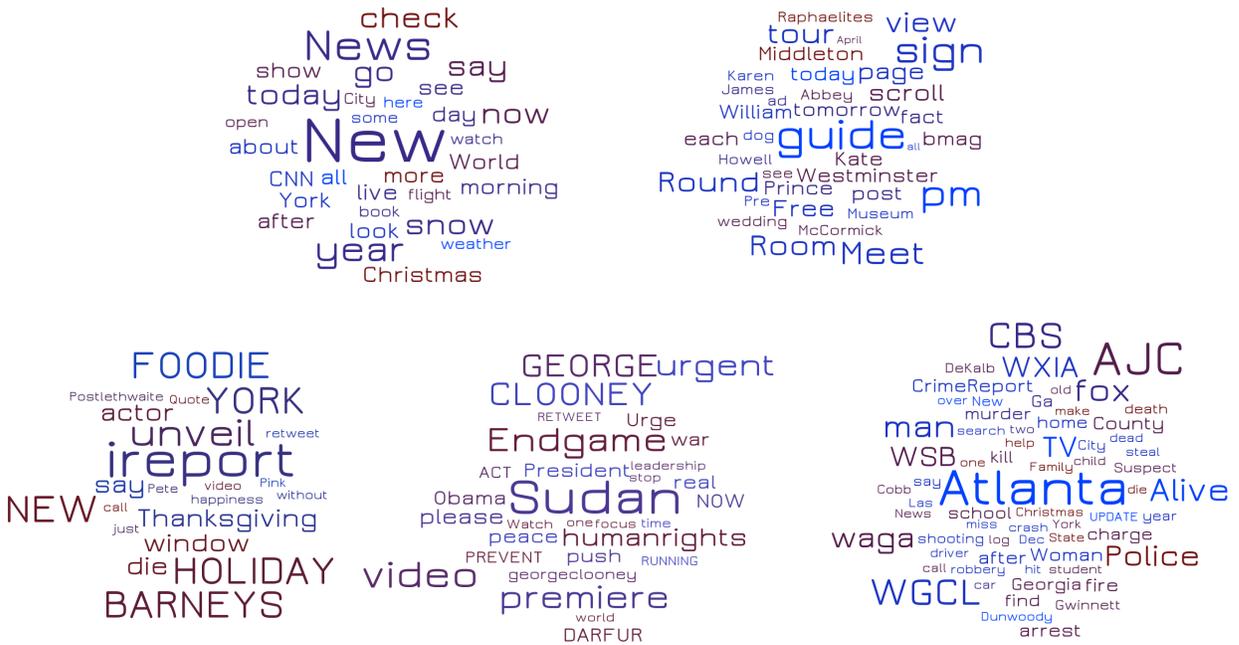

Figure 4. Word cloud representation of some of the clusters found by GSDMM.

As shown in Figure 4, each cluster contains words that relate to an event occurred at around the tweets generation time, according to Section 3.2.3. For instance, the set of terms like "Sudan", "george", "obama", "war" and "Darfur" implies a meeting between George Clooney and Barak Obama about war crimes in Sudan. Or terms such as "Prince", "William", "Kate", "Middleton", "wedding", "April", "Westminster" and "Abbey" indicate Prince William and Kate Middleton's royal wedding on April 2011 at Westminster Abbey. As another example, terms like "AJC[ii]", "Atlanta", "man", "arrest", "shooting" and "police" imply a shooting in Georgia. This sample of five clusters in Figures 4 is shown in order to illustrate what the events literally are in this paper.

## 4.3. Second Component Evaluation

The second component of the proposed framework analyzes the performance of news websites and ranks them according to the measures defined in Section 3.2. The news article dataset used in this paper contains news articles published by CNN, BBC and New York Times. Therefore, the measures are calculated for these three news websites.

In addition to rankings obtained by the defined measures, we obtained another type of ranking, which is independent of the measures and can be used to compare with the rankings obtained from the measures. For this purpose, we employed three search engines and their ranked retrieval results in response to some event-related queries. Since the measure-based rankings are attained by using the events detected from tweets, we built a set of queries using the events such that each query consists of top 10 words with highest weights in the language model of each event. Next, we scored each website based on the ranks of their pages retrieved in response to event-related queries. The detail of our search engine-based website ranking method is shown in Algorithm 1 and the variables used in the Algorithm 1 are defined in Table 3.

```
Algorithm 1 Compute a search engine based ranking of news websites using events
Require: events ē, events' language models Θ⃗, websites w⃗, search engine s
    return ranking score of websites, RS⃗(w)
    for all w in w⃗ do
        RS⃗(w) ← 0
    end for
    while there is an element e_i in ē do
        q_i ← getTop10words(θ(e_i))
        r⃗_i ← getTop10RetrievalResults(s, q_i)
        for every result r_i^j, in retrieval results r⃗_i do
            v_i^j ← getSourceWebsite(r_i^j)
            Rank_i^j ← getRank(r_i^j)
            for every website w_i in w⃗ do
                if v_i^j equals w_i then
                    RS(w_i) ← RS(w_i) + (10 − Rank_i^j)
                    break
                else
                    continue
                end if
            end for
        end for
    end while
```

Table 3. Notations.

| Definition | Variable |
| --- | --- |
| Events detected from tweets | $\vec{e}$ |
| Language model of event $e_i$ | $\theta(e_i)$ |
| Under study news websites | $\vec{w}$ |
| Search engine | s |
| Ranking Score for website y | RS(y) |
| $j^{th}$ document in the retrieval results of query $q_i$ | $r_i^j$ |

In this paper, we chose three prominent search engines including Google, Yahoo and Bing and used the event-related queries to extract search engine-based rankings of the news websites. Table 4 and Table 5 contain the measure-based and search engine-based scores and rankings of the news websites respectively.

Table 4. Measure-based rankings of the news websites.

| Rank | Measures | | | | | |
|---|---|---|---|---|---|---|
| | Diversity | | News_Completeness | | Speed | |
| | News website | score | News website | score | News website | score |
| 1 | NYTimes | 0.450 | BBC | -0.279 | BBC | 2.554 |
| 2 | BBC | 0.333 | CNN | -0.276 | NYTimes | 1.842 |
| 3 | CNN | 0.156 | NYTimes | -0.283 | CNN | 1.098 |

Table 5. Search engine-based rankings of the news websites.

| Rank | Search Engine | | | | | |
|---|---|---|---|---|---|---|
| | Google Ranking | | Yahoo Ranking | | Bing Ranking | |
| | News website | score | News website | score | News website | score |
| 1 | CNN | 4 | - | 0 | CNN | 20 |
| 2 | NYTimes | 3 | - | 0 | NYTimes | 18 |
| 3 | BBC | 0 | - | 0 | BBC | 0 |

As shown in Table 4, the highest diversity value belongs to NYTimes and CNN has the lowest performance in covering different events. The news completeness scores show that all three websites perform almost the same in reporting the details of each event and in terms of speed, BBC outperforms NYTimes and CNN which implies its faster performance in reporting events (of the dataset) in comparison with NYTimes and CNN. Each of the news websites rankings in Table 4, obtained by calculating one of the proposed measures for all news websites can benefit consumers in many ways. The ranking of news websites can act as a standalone application and provide the news consumers an up-to-date ranking of news websites based on the viewpoint of general public about three aspects of news content. These aspects are:

- the set of important events: which is derived from abundance of tweets about those events and is obtained by the event detection method in the proposed framework,
- the description of each event: which is derived from people's narratives of events in their tweets and is represented using language modeling approach in the proposed framework
- the time that each event has occurred: based on the publish time of the burst of tweets about the event

To be more specific, the ranking of news websites based on speed measure benefits the news consumers who wish to be notified of breaking news immediately. This kind of users prefer to read the publications of top ranked websites according to the speed measure. The ranking of news websites based on news completeness measure benefits the users who prefer to find and read the most complete and detailed reports about each event. Similarly, the news diversity measure ranks the news websites according to the preference of news consumers who like to know something about all of the events.

The analysis results can also benefit other news-related tasks such as news recommendation and news retrieval. For example, the ranking based on the news diversity measure can be utilized in finding more candidates to be recommended to user in news recommendation task and consequently improve the recall value in this task. Moreover, the news completeness measure can be effectively used in news recommendation task to adjust the recommended news articles based on their publishing websites' rankings. The news retrieval is the other task that can benefit from news websites analysis results by using the ranking scores of news websites to better estimate the scores of retrieval results. As shown in Table 4, the websites' rankings based on different measures are not necessarily the same, as in some cases, being the first website to report an event, i.e. obtaining higher speed measure value, leads to publishing less comprehensive and detailed article about the event. Therefore, each single measure should not be used for representing the whole performance of a website comprehensively.

Results in Table 5 show that search engine-based rankings of CNN, NYTimes and BBC by Google and Bing are the same although their scores are different. It can implicitly indicate that these two search engines have similar evaluation of the news websites regarding the set of events detected from tweets since their retrieval results in response to the event-related queries were similar. However, it cannot be generalized as a statement about all search engines as the ranking scores calculated from Yahoo results are zero since none of its top 10 retrieval results in response to the event-related queries are from any of the news websites being studied here.

Comparing the search engine-based rankings with the measure-based ranking (Table 4 and Table 5) shows that in general they are not ranking the news websites the same. This can be due to the fact that the search engine-based ranking of the news websites is a secondary output of the search engine. In other words, since the news website rankings are achieved from the retrieval results of the search engines, it can be affected by retrieval process considerations which is not the same problem as news website ranking problem. By considering the Twitter users as a sample of people and the measure-based rankings (rankings obtained from the proposed framework) as a ranking derived from the people's viewpoint, the difference between the rankings of Table 4 and Table 5 imply that rankings of search engines do not necessarily comply with the people's point of view.

## 5. Conclusions and Future Works

In this paper, we proposed a framework that employs user generated content to analyze and rank news websites automatically. This framework consists of two components: event detection component and news websites ranking component. The first component uses tweets written by Twitter news publisher users to detect events. The extracted events are used by the second component, which ranks online news websites using three measures. The measures are defined to assess the diversity of events reported by online news websites, the completeness of news articles reporting an event, and the speed of online news websites in reporting the events. The performance of the proposed framework depends on the performance of its constituent components so the imperfect precision in each of the components such as imperfect precision of news publisher detection or event detection impacts the final results of the framework. We employed the proposed framework to rank three well-known news websites including BBC, CNN and NYTimes. The results show that NYTimes has the best performance in terms of diversity measure, BBC has the highest speed score and the news completeness measure for all three news websites are nearly the same.

One of the future directions to explore is to expand the set of ranking measures to assess more aspects of an online news website's performance. For example, a metric that gauges the validity of the events published by the website through counting the number of rumors reported by the website, a metric that measures the number of events reported by the website exclusively or a metric that measures how briefly the events are reported by the websites can improve the performance analysis of news websites. Moreover, measuring the subjectivity level of textual content published by news websites using subjectivity detection methods such as [80] can be another future work. Another approach to continue this study is applying the proposed framework to analyze the performance of more news websites, specially a mixture of local and global news websites to analyze and compare their performance in reporting local and global events separately. Applying a questionnaire-based website evaluation method to rank the same set of news websites and compare the results with the ranking results of the proposed framework can be another future work. Employing the results of the proposed framework in other new-related tasks such as news recommendation and news retrieval to improve their results by using the rankings of news websites can be another approach to extend this work.

**Notes**
  I. polarity
 II. AJC (Atlanta Journal-Constitution) is the name of a daily newspaper in Atlanta

# References

[1] Bar-Ilan J, Zhu Z and Levene M. Topic-specific analysis of search queries. In: *Proceedings of the 2009 workshop on Web Search Click Data*, 2009 Feb 9, pp. 35-42. ACM.

[2] Pew Research Center, http://www.pewresearch.org/ (accessed 21 December 2018).

[3] The Evolving Role of News on Twitter and Facebook, Pew Research Center's Journalism Project, http://www.journalism.org/2015/07/14/the-evolving-role-of-news-on-twitter-and-facebook/ (accessed 21 December 2018).

[4] Holton AE, Coddington M and Gil de Zúñiga H. Whose news? Whose values? Citizen journalism and journalistic values through the lens of content creators and consumers. *Journalism Practice*. 2013 Dec 1; 7(6): 720-737.

[5] Al-Radaideh QA, Abu-Shanab E, Hamam S and Abu-Salem H. Usability Evaluation of Online News Websites: A User Perspective Approach. *International Journal of Human and Social Sciences*. 2011; 6(2): 114-122.

[6] Bucy EP. Media credibility reconsidered: Synergy effects between on-air and online news. *Journalism & Mass Communication Quarterly*. 2003 Jun; 80(2): 247-264.

[7] Chung CJ, Nam Y and Stefanone MA. Exploring online news credibility: The relative influence of traditional and technological factors. *Journal of Computer-Mediated Communication.* 2012 Jan 1; 17(2): 171-186.

[8] Hong S and Kim J. Architectural criteria for website evaluation–conceptual framework and empirical validation. *Behavior & Information Technology.* 2004 Sep 1; 23(5): 337-357.

[9] Kuan-Tsae H, Lee Yang W and Wang Richard Y. *Quality information and knowledge.* New Jersey: Prentice Hall, 1999.

[10] Kirakowski J. Questionnaires in usability engineering, A List of Frequently Asked Questions, Web-site compiled by: Jurek Human Factors Research Group, Cork, Ireland, 2006.

[11] Kruikemeier S and Lecheler S. News consumer perceptions of new journalistic sourcing techniques. *Journalism Studies.* 2018 Apr 4; 19(5): 632-649.

[12] Kirchner M. Evaluation, repair, and transformation of Web pages for Web content accessibility. Review of some available tools. In: *Proceedings 4th International Workshop on Web Site Evolution.* IEEE Computer Society Press: Los Alamitos CA, 2002; pp. 65–72.

[13] Schubert P and Selz D. Web assessment-a model for the evaluation and the assessment of successful electronic commerce applications. In: *Proceedings of the Thirty-First Hawaii International Conference on System Sciences,* 1998 Jan 6, Vol. 4, pp. 222-231. IEEE.

[14] Ivory MY and Hearst MA. The state of the art in automating usability evaluation of user interfaces. *ACM Computing Surveys (CSUR)* 2001 Dec 1; 33(4): 470-516.

[15] Smith AG. Testing the surf: criteria for evaluating Internet information resources. *Public Access-Computer Systems Review.* 1997 Jul; 8(3).

[16] Brajnik G. Using automatic tools in accessibility and usability assurance processes. In: *ERCIM Workshop on User Interfaces for All*, 2004 Jun 28, pp. 219-234, Springer, Berlin, Heidelberg.

[17] Nielsen J. Usability inspection methods. In: *Conference companion on Human factors in computing systems*, 1994 Apr 28, pp. 413-414. ACM.

[18] Pjesivac I, Geidner N and Cameron J. Social credibility online: The role of online comments in assessing news article credibility. *Newspaper Research Journal.* 2018 Mar; 39(1): 18-31.

[19] Treiblmaier H and Pinterits A. Developing metrics for web sites. *Journal of Computer Information Systems.* 2010 Mar 1; 50(3): 1-10.

[20] Cassidy WP. Online news credibility: An examination of the perceptions of newspaper journalists. *Journal of Computer-Mediated Communication.* 2007 Jan 1; 12(2): 478-498.

[21] Havlícek Z, Lohr V, Šmejkalová M, Grosz J and Benda P. Agritourism Farms-Evaluation of Their Websites Quality and Web 2.0. *AGRIS on-line Papers in Economics and Informatics.* 2013 Mar 1; 5(1): 31.

[22] Law R, Qi S and Buhalis D. Progress in tourism management: A review of website evaluation in tourism research. *Tourism management.* 2010 Jun 1; 31(3): 297-313.

[23] Ebenezer C. Usability evaluation of an NHS library website. *Health Information & Libraries Journal.* 2003 Sep; 20(3): 134-142.

[24] Li X. Web page design and graphic use of three US newspapers. *Journalism & Mass Communication Quarterly.* 1998 Jun; 75(2): 353-365.

[25] Abdullah R and Wei KT. Usability measurement of Malaysia online news websites. *International Journal of Computer Science and Network Security.* 2008 May; 8(5): 159-165.

**Under Review**

[26] Zhitomirsky-Geffet M, David E, Koppel M and Uzan H. Utilizing overtly political texts for fully automatic evaluation of political leaning of online news websites. *Online Information Review.* 2016 Jun 13; 40(3): 362-379.

[27] Fernandez A, Insfran E and Abrahão S. Usability evaluation methods for the web: A systematic mapping study. *Information and software Technology.* 2011 Aug 1; 53(8): 789-817.

[28] Dominic PD, Jati H and Kannabiran G. Performance evaluation on quality of Asian e-government websites–an AHP approach. *International Journal of Business Information Systems.* 2010 Jan 1; 6(2): 219-239.

[29] Rodríguez RA, Estévez EC, Giulianelli DA and Vera PM. Assessing e-governance maturity through municipal websites–measurement framework and survey results. In: *Proceedings of the 6th workshop on software engineering*, Argentinean computer science conference San Salvador de Jujuy, Argentina, October 5–9.

[30] Yan P, Zhang Z and Garcia R. Automatic website comprehensibility evaluation. In: *Proceedings of the IEEE/WIC/ACM international Conference on Web intelligence*, 2007 Nov 2, pp. 191-197. IEEE Computer Society.

[31] Zahran DI, Al-Nuaim HA, Rutter MJ and Benyon D. A comparative approach to web evaluation and website evaluation methods. *International Journal of Public Information Systems.* 2014 Jun 3; 10(1).

[32] Chiou WC, Lin CC and Perng C. A strategic framework for website evaluation based on a review of the literature from 1995–2006. *Information & management.* 2010 Aug 1; 47(5-6): 282-290.

[33] Tsai P. A survey of empirical usability evaluation methods. *GSLIS Independent Study.* 2006:1-8.

[34] Zhang P, Von Dran GM. Satisfiers and dissatisfiers: A two-factor model for website design and evaluation. *Journal of the American society for information science.* 2000; 51(14): 1253-1268.

[35] Bakker T, Trilling D, de Vreese C, Helfer L and Schönbach K. The context of content: the impact of source and setting on the credibility of news. *Recherches Commun.* 2013 Dec 13; 40(40): 151-168.

[36] Broersma M and Graham T. Twitter as a news source: How Dutch and British newspapers used tweets in their news coverage, 2007–2011. *Journalism practice.* 2013 Aug 1; 7(4): 446-464.

[37] Hermida A. # Journalism: Reconfiguring journalism research about Twitter, one tweet at a time. *Digital journalism.* 2013 Oct 1; 1(3): 295-313.

[38] Hermida A, Fletcher F, Korell D and Logan D. Share, like, recommend: Decoding the social media news consumer. *Journalism Studies.* 2012 Oct 1; 13(5-6): 815-824.

[39] Lecheler S and Kruikemeier S. Re-evaluating journalistic routines in a digital age: A review of research on the use of online sources. *New media & society.* 2016 Jan; 18(1): 156-171.

[40] Moon SJ, Hadley P. Routinizing a new technology in the newsroom: Twitter as a news source in mainstream media. *Journal of Broadcasting & Electronic Media* 2014 Apr 3; 58(2): 289-305.

[41] Allan J, editor. *Topic detection and tracking: event-based information organization.* Springer Science & Business Media. 2012 Dec 6.

[42] Ahn D. The stages of event extraction. In: *Proceedings of the Workshop on Annotating and Reasoning about Time and Events*, 2006 Jul 23, pp. 1-8. Association for Computational Linguistics.

[43] Allan J, Papka R and Lavrenko V. On-line new event detection and tracking. In: *Proceedings of the 21st annual international ACM SIGIR conference on Research and development in information retrieval*, 1998 Aug 1, pp. 37-45. ACM.

[44] Hatzivassiloglou V and Filatova E. Domain-independent detection, extraction, and labeling of atomic events. In: *Proceedings of Recent Advances in Natural Language Processing (RANLP '03)*, 2003, pp. 145-152.

[45] Hatzivassiloglou V, Gravano L and Maganti A. An investigation of linguistic features and clustering algorithms for topical document clustering. In: *Proceedings of the 23rd annual international ACM SIGIR conference on Research and development in information retrieval*, 2000 Jul 1, pp. 224-231. ACM.

[46] Ji H and Grishman R. Refining event extraction through cross-document inference. In: *Proceedings of ACL-08: HLT*, 2008, pp. 254-262.

[47] Kumaran G and Allan J. Text classification and named entities for new event detection. In: *Proceedings of the 27th annual international ACM SIGIR conference on Research and development in information retrieval*, 2004 Jul 25, pp. 297-304. ACM.

[48] Makkonen J, Ahonen-Myka H and Salmenkivi M. Simple semantics in topic detection and tracking. *Information retrieval*. 2004 Sep 1; 7(3-4): 347-368.

[49] Yang Y, Pierce T and Carbonell J. A study of retrospective and on-line event detection. In: *Proceedings of the 21st annual international ACM SIGIR conference on Research and development in information retrieval*, 1998 Aug 1, pp. 28-36. ACM.

[50] Zhang K, Zi J and Wu LG. New event detection based on indexing-tree and named entity. In: *Proceedings of the 30th annual international ACM SIGIR conference on Research and development in information retrieval*, 2007 Jul 23, pp. 215-222. ACM.

[51] Becker H, Naaman M and Gravano L. Event Identification in Social Media. In: *International Workshop on the Web and Databases (WebDB)*. 2009 Jun 28.

[52] Becker H, Naaman M and Gravano L. Selecting Quality Twitter Content for Events. In: *ICWSM*, 2011 Jul 17;11.

[53] Van Canneyt S, Feys M, Schockaert S, Demeester T, Develder C and Dhoedt B. Detecting newsworthy topics in twitter. In: *Proceedings of SNOW 2014 Data Challenge*, 2014, pp. 1-8.

[54] Chen Y, Amiri H, Li Z and Chua TS. Emerging topic detection for organizations from microblogs. In: *Proceedings of the 36th international ACM SIGIR conference on Research and development in information retrieval*, 2013 Jul 28, pp. 43-52. ACM.

[55] Cordeiro M. Twitter event detection: combining wavelet analysis and topic inference summarization. In: *Doctoral symposium on informatics engineering*. 2012 Jan 26, pp. 11-16.

[56] Li C, Sun A and Datta A. Twevent: Segment-based event detection from tweets. In: *Proceedings of the 21st ACM international conference on Information and knowledge management*, 2012 Oct 29, pp. 155-164. ACM.

[57] Paltoglou G. Sentiment-based event detection in Twitter. *Journal of the Association for Information Science and Technology*. 2016 Jul; 67(7): 1576-1587.

[58] Weng J and Lee BS. Event detection in Twitter. In: *ICWSM*, 2011 Jul 17 ;11: 401-408.

[59] Abel F, Gao Q, Houben GJ and Tao K. Analyzing user modeling on Twitter for personalized news recommendations. In: *International Conference on User Modeling, Adaptation, and Personalization*, 2011 Jul 11, pp. 1-12. Springer, Berlin, Heidelberg.

**Under Review**

[60] Abel F, Gao Q, Houben GJ, Tao K. Semantic enrichment of Twitter posts for user profile construction on the social web. In: *Extended semantic web conference*. 2011 May 29, pp. 375-389. Springer, Berlin, Heidelberg.

[61] Gao W, Li P and Darwish K. Joint topic modeling for event summarization across news and social media streams. In: *Proceedings of the 21st ACM international conference on Information and knowledge management*. 2012 Oct 29, pp. 1173-1182. ACM.

[62] Lourentzou I, Dyer G, Sharma A and Zhai C. Hotspots of news articles: Joint mining of news text & social media to discover controversial points in news. In: *International Conference on Big Data (Big Data)*. 2015 Oct 1, pp. 2948-2950. IEEE.

[63] Popescu AM and Pennacchiotti M. Detecting controversial events from Twitter. In: *Proceedings of the 19th ACM international conference on Information and knowledge management*, 2010 Oct 26, pp. 1873-1876. ACM.

[64] Brants T, Chen F and Farahat A. A system for new event detection. In: *Proceedings of the 26th annual international ACM SIGIR conference on Research and development in information retrieval*, 2003 Jul 28, pp. 330-337. ACM.

[65] Yang Y, Ault T, Pierce T and Lattimer CW. Improving text categorization methods for event tracking. In: *Proceedings of the 23rd annual international ACM SIGIR conference on Research and development in information retrieval*, 2000 Jul 1, pp. 65-72. ACM.

[66] Yang Y, Carbonell JG, Brown RD, Pierce T, Archibald BT and Liu X. Learning approaches for detecting and tracking news events. *IEEE Intelligent Systems and Their Applications*. 1999 Jul 1; 14(4): 32-43.

[67] Fung GP, Yu JX, Yu PS and Lu H. Parameter free bursty events detection in text streams. In: *Proceedings of the 31st international conference on Very large data bases*, 2005 Aug 30, pp. 181-192. VLDB Endowment.

[68] Goorha S and Ungar L. Discovery of significant emerging trends. In: *Proceedings of the 16th ACM SIGKDD international conference on Knowledge discovery and data mining*, 2010 Jul 25, pp. 57-64. ACM.

[69] Kleinberg J. Bursty and hierarchical structure in streams. *Data Mining and Knowledge Discovery.* 2003 Oct 1; 7(4): 373-397.

[70] Garg M and Kumar M. Review on event detection techniques in social multimedia. *Online Information Review.* 2016 Jun 13; 40(3): 347-361.

[71] Krestel R, Werkmeister T, Wiradarma TP and Kasneci G. Tweet-recommender: Finding relevant tweets for news articles. In: *Proceedings of the 24th International Conference on World Wide Web*. 2015 May 18, pp. 53-54. ACM.

[72] Shi B, Ifrim G and Hurley N. Be in the know: Connecting news articles to relevant Twitter conversations. arXiv preprint arXiv:1405.3117. 2014 May 13.

[73] Cooper GF and Herskovits E. A Bayesian method for the induction of probabilistic networks from data. *Machine learning.* 1992 Oct 1; 9(4): 309-347.

[74] Hall M, Frank E, Holmes G, Pfahringer B, Reutemann P and Witten IH. The WEKA data mining software: an update. *ACM SIGKDD explorations newsletter*. 2009 Nov 16; 11(1): 10-18.

[75] Yin J and Wang J. A Dirichlet multinomial mixture model-based approach for short text clustering. In: *Proceedings of the 20th ACM SIGKDD international conference on Knowledge discovery and data mining*, 2014 Aug 24, pp. 233-242. ACM.

[76] Fuglede B and Topsoe F. Jensen-Shannon divergence and Hilbert space embedding. In: *Proceedings of the IEEE International Symposium on Information Theory*, 2004 Jun 27, p. 31.

[77] Analyzing User Modeling on Twitter for Personalized News Recommendations, http://wis.ewi.tudelft.nl/umap2011/ (accessed 21 December 2018)

[78] Manning C, Surdeanu M, Bauer J, Finkel J, Bethard S and McClosky D. The Stanford CoreNLP natural language processing toolkit. In: *Proceedings of 52nd annual meeting of the association for computational linguistics: system demonstrations*, 2014, pp. 55-60.

[79] Friedman N, Geiger D and Goldszmidt M. Bayesian network classifiers. *Machine learning.* 1997 Nov 1; 29(2-3): 131-163.

[80] Karimi S and Shakery A. A language-model-based approach for subjectivity detection. *Journal of Information Science.* 2017 Jun; 43(3): 356-377.

[81] Broersma M and Graham T. Social media as beat: Tweets as a news source during the 2010 British and Dutch elections. *Journalism Practice.* 2012 Jun 1; 6(3): 403-419.

[82] Abanumy A, Al-Badi A and Mayhew P. e-Government Website accessibility: In-depth evaluation of Saudi Arabia and Oman. *The Electronic Journal of e-government.* 2005; 3(3): 99-106.

[83] Elling S, Lentz L and De Jong M. Website evaluation questionnaire: Development of a research-based tool for evaluating informational websites. In: *International Conference on Electronic Government*, 2007 Sep 3, pp. 293-304. Springer, Berlin, Heidelberg.

[84] Hasan M, Orgun MA and Schwitter R. A survey on real-time event detection from the Twitter data stream. *Journal of Information Science.* 2017 March.

[85] Seifikar M and Farzi S. A comprehensive study of online event tracking algorithms in social networks. *Journal of Information Science.* 2018 Jul 3:0165551518785548.

[86] Farzindar A and Khreich W. A survey of techniques for event detection in Twitter. *Computational Intelligence.* 2015; 31(1): 132-164.

[87] Demirsoz O and Ozcan R. Classification of news-related tweets. *Journal of Information Science.* 2017 Aug; 43(4): 509-524.

[88] Fang Y, Zhang H, Ye Y and Li X. Detecting hot topics from Twitter: A multiview approach. *Journal of Information Science.* 2014 Oct; 40(5): 578-593.

[89] Lee CH and Chien TF. Leveraging microblogging big data with a modified density-based clustering approach for event awareness and topic ranking. *Journal of Information Science.* 2013 Aug; 39(4): 523-543.

[90] Shapira O, Ronen H, Adler M, Amsterdamer Y, Bar-Ilan J and Dagan I. Interactive abstractive summarization for event news tweets. In: *Proceedings of the Conference on Empirical Methods in Natural Language Processing: System Demonstrations*, Sep. 2017, pp. 109–114.

[91] Zhai C. Statistical language models for information retrieval. Synthesis Lectures on Human Language Technologies. 2008 Dec 5;1(1):1-41.

[92] Mele I, Bahrainian SA and Crestani F. Event mining and timeliness analysis from heterogeneous news streams. *Information Processing & Management.* May 2019, 56(3): 969-993.

[93] Verma, R, Karimi, S, Lee, D, Gnawali, O and Shakery, A. Newswire versus Social Media for Disaster Response and Recovery. arXiv e-prints, arXiv:1906.10607. 2019.